\documentclass[12pt]{article}
\usepackage{graphicx}
\usepackage[compress]{cite}

\newcommand\pubnumber{DPF2015-192}
\newcommand\pubdate{August 17, 2015}

\def\iit{Department of Physics\\ Illinois Institute of Technology\\ Chicago,
Illinois 60616-3793, USA}
\def\fsu{Physics Department\\ Florida State University\\ Tallahassee, Florida 32306-4350, USA}

\def\support{\footnote{Presenting author. zack.sullivan@iit.edu}}

\def\Title#1{\begin{center} {\Large #1 } \end{center}}
\def\Author#1{\begin{center}{ \sc #1} \end{center}}
\def\Address#1{\begin{center}{ \it #1} \end{center}}

\newcommand\pubblock{\rightline{\begin{tabular}{l} \pubnumber\\
         \pubdate  \end{tabular}}}
\newenvironment{Abstract}{\begin{quotation}  }{\end{quotation}}
\newenvironment{Presented}{\begin{quotation} \begin{center} 
             PRESENTED AT\end{center}\bigskip 
      \begin{center}\begin{large}}{\end{large}\end{center} \end{quotation}}
\def\Acknowledgments{\bigskip  \bigskip \begin{center} \begin{large}
             \bf ACKNOWLEDGMENTS \end{large}\end{center}}
\def\beq{\begin{equation}}
\def\eeq#1{\label{#1}\end{equation}}
\def\eeqn{\end{equation}}

\def\beqa{\begin{eqnarray}}
\def\eeqa#1{\label{#1}\end{eqnarray}}
\def\eeqan{\end{eqnarray}}

\let\bar=\overbar

\def\Dslash{\not{\hbox{\kern-4pt $D$}}}
\def\dslash{\not{\hbox{\kern-2pt $\del$}}}

\def\msb{{\bar{\ssstyle M \kern -1pt S}}}

\begin{document}
\begin{titlepage}
\pubblock

\vfill
\Title{Improving parton distribution uncertainties\\ in a $W$ mass measurement
at the LHC}
\vfill
\Author{Zack Sullivan\support}
\Address{\iit}
\Author{Seth Quackenbush}
\Address{\fsu}
\vfill
\begin{Abstract}
We reexamine the dominant contribution of parton distribution function
(PDF) uncertainties to the $W$ mass measurement, and determine their
contribution is $\pm 39(30)$ MeV when running the Large Hadron
Collider at 7(13) TeV.  We find that spurious correlations in older
PDF sets led to over-optimistic assumptions regarding normalization to
$Z$ observables.  In order to understand the origin of the large
uncertainties we break down the contribution of the PDF errors into
effects at the hard matrix element level, in showering, and in
sensitivity to finite detector resolutions.

Using CT10, CT10W, and charm enhanced PDF sets in comparison to older
PDF sets, we develop a robust analysis that examines correlations
between transverse mass reconstructions of $W$ and $Z$ decays (scaled by
cos $\theta_W$) to leptons.  We find that central leptons ($|\eta_l|
< 1.3$) from $W$ and $Z$ bosons carry the most weight in reducing the
PDF uncertainty, and estimate a PDF error of $^{+10}_{-12}$ MeV is
achievable in a $W$ mass measurement at the LHC.  Further reductions of
the $W$ mass uncertainty will require improved fits to the parton
distribution functions.
\end{Abstract}
\vfill
\begin{Presented}
DPF 2015\\
The Meeting of the American Physical Society\\
Division of Particles and Fields\\
Ann Arbor, Michigan, August 4--8, 2015\\
\end{Presented}
\vfill
\end{titlepage}
\def\thefootnote{\fnsymbol{footnote}}
\setcounter{footnote}{0}

\section{Introduction}

\indent\indent While the mass of the $W$ boson is currently known to
an astonishing two parts in ten thousand, the uncertainty is dominated
by the theoretical estimate of parton distribution function (PDF)
errors.  The combined efforts of the CDF and D0 collaborations at the
Fermilab Tevatron have established $M_W=80.385\pm 15$ GeV
\cite{Aaltonen:2013iut}.  Despite this accomplishment, the recent
discovery of a Higgs-like boson \cite{Aad:2012tfa,Chatrchyan:2012ufa}
shows the most sensitive constraint on consistency of the Standard
Model global fit \cite{Agashe:2014kda} is the $W$ mass uncertainty.
In addition, most models of enhanced symmetry predict significant
quantum corrections to the $W$ mass, e.g., supersymmetry can shift the
mass by 2--20 MeV \cite{Heinemeyer:2006px,Baak:2013fwa}, hence a
measurement of the mass to 5 MeV or better at the LHC is desirable
\cite{Baak:2013fwa}.

In a recently published article \cite{Quackenbush:2015yra}, we
demonstrate that current PDF uncertainties would lead to a $\pm
39(30)$~MeV error in a LHC measurement at 7(13) TeV.  However, we
propose a new technique to reduce that uncertainty to $\pm 10$--12 MeV
or better that is robust against spurious correlations that plague
earlier theoretical error estimates.  Hence, with improvements to
individual PDF uncertainties in combination with our method, a 5 MeV
measurement of $M_W$ should be achievable using data from the LHC.

\section{PDF contribution to the $M_W$ uncertainty}

\indent\indent The $W$ mass is measured by fitting templates to
leptonic observables from $W$ boson decay to $e\nu_e$ or $\mu\nu_\mu$.
The most reliably predicted observable is the transverse mass variable
\begin{equation} 
M_T = \sqrt{2p_T^l E_T^\mathrm{miss} (1-\cos(\Delta\phi_{l,\mathrm{miss}}))}.
\end{equation}
While predictions by RESBOS \cite{Ladinsky:1993zn,Landry:2002ix} and
measurements \cite{Aaltonen:2013vwa} of lepton transverse momentum
$P_T^l$ or missing energy $E_T^\mathrm{miss}$ are individually good,
they are both subject to errors in the estimate of $W$ boson recoil
from soft QCD radiation.  These errors are suppressed in $M_T$.  In
addition, PDF errors turn out to be significantly larger for $p_T^l$
and $E_T^\mathrm{miss}$ alone.  Hence, we concentrate on $M_T$ below.

The first key observation of Ref.\ \cite{Quackenbush:2015yra} is that
PDF errors enter in multiple places into the prediction of the $W$
mass templates.  Hard matrix-element calculations show that PDFs shift
the rapidity distribution of the $W$ boson, causing acceptance
uncertainties in the fits.  We find a larger uncertainty arises from
soft QCD initial state radiation.  This latter effect probes a
higher-$x$, lower-$Q^2$ region of the PDFs than the matrix element ---
a more poorly constrained region.  Finally, the $M_T$ calculation is
afflicted by small additional shifts caused by finite detector
resolution providing access to a broader, less well-constrained region
of the PDFs.  This decomposition is critical, as it points to where
progress can be made in improving PDF uncertainties.

Older estimates of PDF uncertainties at the LHC were large, $\pm
25$~MeV \cite{Buge:2006dv,Besson:2008zs}, though recent predictions
that missed some of the effects above claim that they are under
control at the $\pm 10$~MeV level
\cite{Bozzi:2011ww,Rojo:2013nia,Bozzi:2015hha}.  All other theoretical
and experimental errors are expected to be small \cite{Baak:2013fwa},
and so we concentrate on updating the PDF errors using modern CTEQ
CT10 and CT10W \cite{Lai:2010vv} PDFs.  We find the uncertainties in
Table \ref{tab:errors} are significantly larger than previously
estimated.  The increase in error is predominantly due to the removal
of a restriction on the form of the $s$ quark PDF in newer PDF sets
that was causing spurious correlations in the error estimates.
Another large uncertainty arises due to $u/d$ valence discrepancies in
the forward detector region at the Tevatron that led to the need for
separate CT10 and CT10W fits.  The 10 MeV difference between the error
estimates should be taken as an unknown systematic in the PDF fits
that will be improved resolved with data from the LHC.  Finally, we
point out that the charm PDF is still somewhat restricted in form in
the conventional PDF fits, and we estimate\footnote{Our estimate uses
  the CTEQ 6.6C2 \protect\cite{Nadolsky:2008zw} intrinsic charm PDF to
  bound the error.}  that up to an additional 10 MeV uncertainty can
arise in more flexible charm fits.

\begin{table}[htbp]
\begin{center}
\begin{tabular}{ c | c c c | c c c }
 & & 7 TeV & & & 13 TeV &\\
& CT10 & CT10W & CT10+IC & CT10 & CT10W & CT10+IC \\ \hline
$m_T$ error & $^{+39}_{-39}$ & $^{+27}_{-27}$ & $^{+39}_{-40}$ & $^{+30}_{-27}$ & 
$^{+25}_{-24}$ & $^{+30}_{-31}$ \\ \hline
$p_T^e$ error & $^{+59}_{-54}$ & $^{+46}_{-45}$ & $^{+59}_{-65}$ & $^{+54}_{-52}$ &
$^{+48}_{-50}$ & $^{+54}_{-65}$\\ \hline
\end{tabular}
\caption{Updated uncertainties in MeV at the LHC from various CTEQ
  PDFs for $M_T$ and $P_T^l$ measurements.  IC estimates an additional
  unknown charm quark contribution to the PDF error normally left out
  of the standard Hessian calculation.}
\label{tab:errors}
\end{center}
\end{table}

\section{A robust method for improving PDF errors}

\indent\indent Two types of \textit{in situ} experimental handles
exist to improve the PDF uncertainty contribution to the $W$ mass
measurement. The first handle is the separation of $W$ events into
four categories: $W^+$ vs.\ $W^-$, and central $|\eta_l| < 1.3$ vs.\
forward $|\eta_l|>1.6$ charged leptons.  Attempts to further partition
charged lepton pseudorapidity are marginally effective: valence quark
PDF uncertainties are mostly concentrated in forward events, while sea
quark PDF uncertainties dominate the central region.  Careful account
of correlations between these four regions allows for a reduction in
the PDF uncertainty in the $M_T$ measurement to $\pm 20$ MeV --- an
improvement, but not enough.

The most important improvement to the PDF uncertainties comes in
reducing the PDF contribution to initial state radiation uncertainties
by normalizing the $W$ mass measurement to a measurement of the $Z$
mass, as first suggested in Ref.\ \cite{Giele:1998uh}.  We update this
idea by recommending the $Z$ mass be split into four subsamples:
$Z^\pm$ represents applying identical cuts to the charged lepton from
the $Z$ decay that matches the $W^\pm$ central or forward boson
sample, and basic acceptance cuts are applied to the other lepton.
We then recommend a linear fit to a weighted sum
\begin{equation}
\sum_{i\in W,Z} \alpha_i M_W^i ,
\end{equation}
where $M_W^Z = \cos \theta_W M_Z$, and the $\alpha_i$ are
determined by minimizing the 8x8 correlation matrix formed from $W$
and $Z$ measurements.  For derivations and detailed correlation matrices
see Ref.\ \cite{Quackenbush:2015yra}.

It is important to note that older predictions that attempted to
normalize to the $Z$ measurements suffer from the same false
correlations that afflicted the total PDF error prediction.  E.g.,
correlations that were 85\% before are found to be closer to 40\% in
modern PDFs.  The relevant PDFs are now fairly stable, and so we do
not expect significant further degradation in the correlations.  We
confirm this with our modeling of extreme charm uncertainties.

After minimization we find consistent predictions for the PDF
contribution to the $W$ mass uncertainty shown in Table
\ref{tab:minerr}.  Even after the addition of possible charm
uncertainties, a robust uncertainty of $\pm 10$--12 MeV can be
achieved.  A similar treatment of $p_T^l$ uncertainties estimates a
robust $\pm 18$ MeV uncertainty can be achieved in that channel.

\begin{table}[htbp]
\begin{center}
\begin{tabular}{ c | c c c c}
$M_T^W$  & CT10 & CT10W & CT10+IC & CT10+IC$_\mathrm{opt}$ \\ \hline
7 TeV & $^{+11}_{-10}$ & $^{+8}_{-8}$ & $^{+11}_{-20}$ & $^{+13}_{-12}$ \\\hline
13 TeV & $^{+10}_{-11}$ & $^{+7}_{-11}$ & $^{+10}_{-14}$ & $^{+10}_{-12}$\\\hline
\end{tabular}
\caption{Robust uncertainties in MeV possible at the LHC using the
  proposed cuts and normalization to $Z$ data in a $M_T$ measurement.
  IC estimates greatly reduced charm quark contributions to the PDF
  errors normally left out of the standard Hessian error calculation, and
  can be optimized away.}
\label{tab:minerr}
\end{center}
\end{table}

\section{Conclusions}

\indent\indent The dominant uncertainty in the experimental
measurement of the $W$ boson mass is the theoretical contribution of
PDF errors.  We isolate the origins of these contributions and find
PDFs enter the error estimates in three places: at the matrix element
level (probing medium-$x$, medium $Q^2$ PDFs), in the initial state
QCD radiation (probing large-$xz$, small-$Q^2$ PDFs), and again when
detector effects shift the sensitivity to different PDF regions.

We find that spurious correlations in older PDFs and matrix-element
level analyses led to overoptimistic predictions for PDF uncertainties
in LHC measurements.  However, we introduce a robust method that uses
kinematic features of the data and a correlated relationship to $Z$
initial-state radiation to propose a method that can reach $\pm
10$--12 MeV in a $M_T$ measurement using CT10 PDFs.  With small
improvements to overall PDF fit uncertainties, particularly using
forward $u/d$ data from LHC, we are confident that our method can
reach the desired 5 MeV level of uncertainty and form a rigorous test
of physics beyond the standard model.

\Acknowledgments
This work was supported by the U.S.\ Department of Energy under Award
Nos.\ {DE-SC0008347} and DE-FG02-13ER41942.

\end{document}